# Rate Regions of Secret Key Sharing in a New Source Model[1,2]

Somayeh Salimi[*], Mahmoud Salmasizadeh[†], Mohammad Reza Aref[*]

[*]ISSL Lab., Dept. of Electrical Engineering, Sharif University of Technology, Tehran, Iran

[†]Electronics Research Center, Sharif University of Technology, Tehran, Iran

Email: salimi@ee.sharif.edu, salmasi@sharif.edu, aref@sharif.edu

**Abstract**—A source model for secret key generation between terminals is considered. Two users, namely users 1 and 2, at one side communicate with another user, namely user 3, at the other side via a public channel where three users can observe i.i.d. outputs of correlated sources. Each of users 1 and 2 intends to share a secret key with user 3 where user 1 acts as a wiretapper for user 2 and vice versa. In this model, two situations are considered: communication from users 1 and 2 to user 3 (the forward key strategy) and from user 3 to users 1 and 2 (the backward key strategy). In both situations, the goal is sharing a secret key between user 1 and user 3 while leaking no effective information about that key to user 2, and simultaneously, sharing another secret key between user 2 and user 3 while leaking no effective information about the latter key to user 1. This model is motivated by wireless communications when considering user 3 as a base station and users 1 and 2 as network users. In this paper, for both the forward and backward key strategies, inner and outer bounds of secret key capacity regions are derived. In special situations where one of users 1 and 2 is only interested in wiretapping and not key sharing, our results agree with that of Ahlswede and Csiszar. Also, we investigate some special cases in which the inner bound coincides with the outer bound and secret key capacity region is deduced.

**Keywords**-Information theoretic security, secret key sharing, source model, secret key capacity region.

## I. INTRODUCTION

Because of the open nature of wireless communication networks, sharing secret keys between terminals is a challenging problem. In these environments, terminals have access to common randomness for generating secret keys but the existence of broadcast and multiple access channels in these networks results in unintended information leakage. In this paper, we explore the problem of sharing secret keys between three users who can observe the outputs of some correlated sources. There are two users, namely user 1 and user 2, at one side and another user, namely user 3, at the other side and also public channels between the users. User 1 wishes to share a secret key with user 3 while user 2 acts as a wiretapper and intends to learn information about this key as much as possible. Symmetrically, user 2 wishes to share a secret key with user 3 while user 1 acts as a wiretapper and intends to learn information about this key as much as possible. This model could be realized in wireless environment when user 3 is a base station and users 1 and 2 are curious network users.

[1]Part of this work will be published on Australian Communication Theory Workshop (AusCTW2010) proceeding.
[2] This work was partially supported by Iranian National Science Foundation (INSF) under Contract No. 84.5193.

The rigorous idea of information theoretic security was first introduced by Shannon in [11] where the eavesdropper could listen to all the data transmitted from the transmitter to the receiver. After that, the notion of information theoretic security was characterized by Wyner as the wiretap channel model in which a single source-destination communication link is eavesdropped by a wiretapper via a degraded channel [13]. The secrecy level was measured by equivocation rate at the wiretapper. It was shown in [13] that nonzero secrecy rate can be achieved without using a secret key, if the intended receiver has a communication channel with better quality than the wiretapper. Csiszar and Korner in their seminal work [2] generalized the Wyner's results to less noisy and more capable channels and determined the capacity region of the broadcast channel with confidential message. In [1] and [8], generation of secret key through common randomness was considered by Maurer, Ahlswede and Csiszar. The common randomness can be a source or a channel type. In source common randomness, all terminals including the transmitter, the receiver and the wiretapper could observe i.i.d. outputs of correlated sources. In channel common randomness, there is a noisy broadcast channel from the transmitter to the receiver and the wiretapper. In both the source and channel common randomness, there is a noiseless public channel with unlimited capacity between the transmitter and the receiver where all communication through which can be overheard by the wiretapper. In [1], based on common randomness type, the source and channel models were defined for secret key sharing and in both models, the problem of finding the secret key capacity between the transmitter and the receiver was considered. In the source model, the secret key capacity was characterized when a one-way noiseless public channel with unlimited capacity is available between the transmitter and the receiver. In case a two-way public channel exists between the transmitter and the receiver, the secret key capacity still remains an open problem, however its upper and lower bounds have been improved in [5] and [10]. Secret key generation in a network including more than three terminals has been explored in other works such as [3], [4], [6], [7], [14], [15]. Maurer [9] strengthened the secrecy conditions of [1] and [8] and showed that the results in a weak sense can be established in the strong sense by using the techniques developed in [9].

As mentioned above, the problem of sharing secret keys between terminals which have access to correlated sources was defined in [1], in which the transmitter and the receiver intend to share a key via public channel communications. In this model, a wiretapper who has access to side information correlated with other sources, can listen to the public channel and obtains information about the shared key as much as possible. In this paper, we propose a new model which differs from the source model of [1] (which was described in the previous paragraph), in such a way that both users 1 and 2 attempt to share secret keys with user 3 while user 1 is the wiretapper of user 2's secret key and vice versa. Three users have access to correlated sources and there is a public channel from users 1 and 2 to user 3. To the best of our knowledge, this model has not been investigated so far. For this model, we investigate two situations. In the first, there is a one-way public channel from users 1 and 2 to user 3. This situation is referred to as the forward key strategy and is shown in Fig.1. In the second one, there is a one-way public channel from user 3 to users 1 and 2. This situation is referred to as the backward key strategy and is shown in Fig.2. In both situations, we investigate the inner and outer bounds of the secret key capacity region.

The rest of the paper is organized as follows: in Section II the proposed model and definitions are described. In Section III, related theorems for the upper and lower bounds of the secret key capacity regions are given. Some special cases are considered in Section IV in which the inner bound coincides with the outer bound and the secret key capacity region can be derived. Proofs of the theorems are given in Section V. Conclusion and suggestions for future works are given in Section VI. Some lemmas useful for the proof of theorems are given and proofed in the appendix. Throughout the paper, a random variable is denoted with an upper case letter (e.g $X$ ) and its realization is denoted with the corresponding lower case letter (e.g., $x$ ). We use $X_i^N$ to indicate vector $(X_{i,1}, X_{i,2},..., X_{i,N})$, and use $X_{i,j}^k$ to indicate vector $(X_{i,j}, X_{i,j+1},..., X_{i,k})$ where $i$ denotes the index of the corresponding user.

## II. THE NEW SOURCE MODEL

Users 1, 2 and 3 can, respectively, observe $N$ i.i.d. repetitions of the random variables $X_1, X_2$ and $X_3$. The random variable $X_i$ takes values from the finite set $\mathcal{X}_i$ for $i = 1, 2, 3$. Furthermore, a noiseless public channel with unlimited capacity is available for communication between the three users. User 1 wishes to share a secret key with user 3 while user 2 acts as a wiretapper of user 1's key. Symmetrically and simultaneously, user 2 wishes to share a secret key with user 3 while user 1 acts as a wiretapper of user 2's key. Now, we represent formal definition of the secret key strategy for the new source model.

*Step 0*) Users 1, 2 and 3, respectively, generate random variables $M_1$, $M_2$ and $M_3$ independent of each other such that $M_1, M_2, M_3$ and $(X_1^N, X_2^N, X_3^N)$ are mutually independent. The next steps can be regarded as deterministic.

*Step 1*) At this step, users 1, 2 and 3, respectively, generate $F_{1,1}$, $F_{2,1}$ and $F_{3,1}$ such that $F_{i,1} = f_{i,1}(M_i, X_i^N)$ for $i = 1, 2, 3$ and transmit them over the public channel.

*Steps 2 to k*) At step $j$, user $i$ generates $F_{i,j}$ as a function of $(M_i, X_i^N)$ and the information which has been received from the other users via the public channel. Hence, users 1, 2 and 3, respectively, generate $F_{1,j}$, $F_{2,j}$ and $F_{3,j}$ as functions of the information available at the corresponding user where $F_{1,j} = f_{1,j}(M_1, X_1^N, F_{2,1}^{j-1}, F_{3,1}^{j-1})$, $F_{2,j} = f_{2,j}(M_2, X_2^N, F_{1,1}^{j-1}, F_{3,1}^{j-1})$ and $F_{3,j} = f_{3,j}(M_3, X_3^N, F_{1,1}^{j-1}, F_{2,1}^{j-1})$ , and transmit them over the public channel for $j = 2,...,k$ .

Finally, after $k$ steps, users 1 and 2 compute the keys $K$ and $L$ , respectively, as functions of the information available at each user:

$$K = K(M_1, X_1^N, F_{2,1}^k, F_{3,1}^k) \qquad (1)$$
$$L = L(M_2, X_2^N, F_{1,1}^k, F_{3,1}^k) \qquad (2)$$

and also user 3 computes the keys $\hat{K}$ and $\hat{L}$ as a function of the information available at him:

$$\hat{K} = \hat{K}(M_3, X_3^N, F_{1,1}^k, F_{2,1}^k) \qquad (3)$$
$$\hat{L} = \hat{L}(M_3, X_3^N, F_{1,1}^k, F_{2,1}^k) \qquad (4)$$

where the keys $\hat{K}$ and $\hat{L}$ are intended for sharing as secret keys with users 1 and 2, respectively. The keys $(K, \hat{K})$ and $(L, \hat{L})$ take values from the finite sets $\mathcal{K}$ and $\mathcal{L}$ , respectively.

Now we state the conditions that should be met in the secret key strategy of the described model.

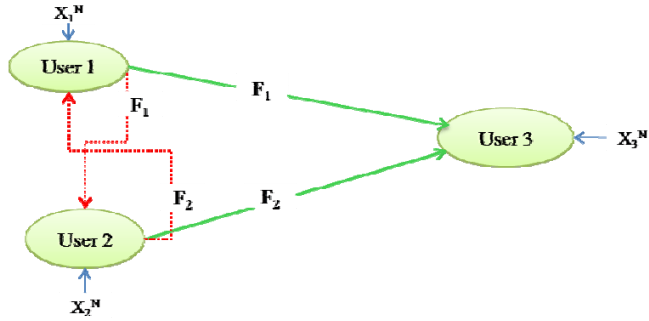

Fig.1. Forward key strategy

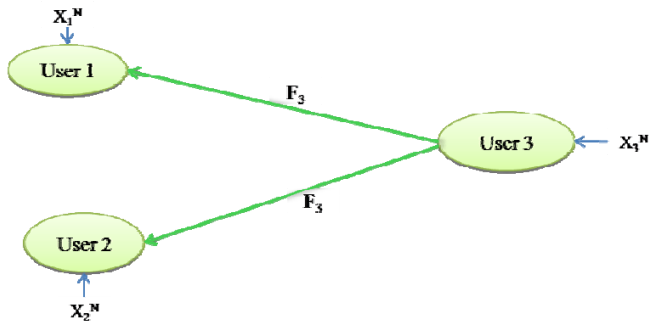

Fig.2. Backward key strategy

***Definition 1***: In the secret key strategy of the source model described above, the secret key rate pair $(R_1, R_2)$ is an achievable rate pair if for every $\varepsilon > 0$ and sufficiently large $N$ , we have:

$$\Pr\{K \neq \hat{K}\} < \varepsilon \text{ and } \Pr\{L \neq \hat{L}\} < \varepsilon \qquad (5)$$

$$\tfrac{1}{N} I(M_2, X_2^N, F_{1,1}^k, F_{3,1}^k; K) < \varepsilon \qquad (6)$$

$$\tfrac{1}{N} I(M_1, X_1^N, F_{2,1}^k, F_{3,1}^k; L) < \varepsilon \qquad (7)$$

$$\tfrac{1}{N} H(K) > R_1 - \varepsilon \text{ and } \tfrac{1}{N} H(L) > R_2 - \varepsilon \qquad (8)$$

$$\tfrac{1}{N} \log |\mathcal{K}| < \tfrac{1}{N} H(K) + \varepsilon \qquad (9)$$

$$\tfrac{1}{N} \log |\mathcal{L}| < \tfrac{1}{N} H(L) + \varepsilon \qquad (10)$$

Equation (5) means that users 1 and 2 can generate secret keys with user 3 and Equations (6) and (7) say that users 1 and 2 have effectively no information about each other's secret key. Equations (9) and (10) are the uniformity conditions for the secret keys.

***Definition 2***: The region containing all the achievable secret key rate pairs $(R_1, R_2)$ is the key capacity region.

In the described model, we consider restricted usage of the public channel, i.e., no more than $k$ usages of the public channel are allowed. In this paper, only the case $k = 1$ is investigated. For this case, when communication is only performed from users 1 and 2 to user 3, forward key capacity region is defined and when communication is only carried out in the reverse direction, i.e., from user 3 to users 1 and 2, backward key capacity region is introduced. We consider both situations in this paper.

### III.  SECRET KEY RATE REGIONS

In this section, we state our main results about the mentioned model.

***Theorem 1 (inner bound of the forward key capacity region)***: In the forward key strategy of the described source model, the rate pair $(R_1, R_2)$ is an achievable key rate pair if:

$$R_1 \geq 0, R_2 \geq 0$$
$$R_1 \leq I(S; X_3 | T, U) - I(S; X_2 | T, U)$$
$$R_2 \leq I(T; X_3 | S, V) - I(T; X_1 | S, V)$$
$$R_1 + R_2 \leq I(S, T; X_3 | U, V) - I(S; X_2 | T, U) - I(T; X_1 | S, V) - I(S; T | U, V)$$

where $U, V, S, T$ are random variables taking values in sufficiently large finite sets and according to the distribution:

$$p(u, v, s, t, x_1, x_2, x_3) = p(u|s) p(v|t) p(s|x_1) p(t|x_2) p(x_1, x_2, x_3).$$

Proof of the achievability is given in Section V. A. However, we explain the intuitive interpretation of Theorem 1. We assume that users 1 and 2 consider the random variables $S$ and $T$ with the distributions $p(s|x_1)$ and $p(t|x_2)$ for sharing keys with user 3, respectively. These random variables should be decoded by user 3 for generating secret keys. To this end, part of information is sent by users 1 and 2 by transmitting realizations of random variables $U$ and $V$ with distributions $p(u|s)$ and $p(v|t)$, respectively. Then, the other part of information should be sent by users 1 and 2 with total rate $H(S,T|U,V,X_3)$, according to the Slepian-Wolf theorem, to enable user 3 to reconstruct $S$ and $T$. Based on the portion of the rate transmitted by each user, there is a tradeoff between the equivocation rates. For justification of the rate $R_1$, we assume that user 1 sends information with the minimum rate $H(S|U,V,X_3,T)$ after sending realizations of $U$. It is obvious that both of the transmissions by user 1 can result in information leakage about $S$ to user 2. The leakage rate would be equal to:

$$I(S;X_2,U) + H(S|U,V,X_3,T)$$

For obtaining $R_1$, we should subtract the leakage rate from $H(S)$ and hence, we have:

$$R_1 = H(S) - I(S;X_2,U) - H(S|U,V,X_3,T) = I(S;U,V,X_3,T) - I(S;X_2,U) \stackrel{(a)}{=} I(S;U,X_3,T) - I(S;X_2,U)$$
$$\stackrel{(b)}{=} I(S;U,X_3,T) - I(S;X_2,U,T) = I(S;X_3|T,U) - I(S;X_2|T,U)$$

where (a) follows from the distribution of $V$ and (b) from the distribution of $T$ that results in $I(S;T|X_2,U) = 0$. Since the minimum rate $H(S|U,V,X_3,T)$ (according to the Slepian–Wolf theorem) is sent by user 1, $R_1$ is smaller than the calculated rate. The same approach can be applied to the rate $R_2$. For the rate $R_1 + R_2$:

$$R_1 + R_2 = H(S) + H(T) - I(S;X_2,U) - I(T;X_1,V) - H(S,T|U,V,X_3)$$
$$= I(S,T;X_3|U,V) - I(S;X_2|T,U) - I(T;X_1|S,V) - I(S;T|U,V)$$

**Theorem 2 (outer bound of the forward key capacity region)**: If the rate pair $(R_1, R_2)$ is an achievable key rate pair in the forward secret key strategy, then it satisfies:

$$R_1 \geq 0, R_2 \geq 0$$
$$R_1 \leq \{I(S;T,X_3|U) - I(S;X_2|U)\}$$
$$R_2 \leq \{I(T;S,X_3|V) - I(T;X_1|V)\}$$

for random variables $U,V,S,T$ which take values in sufficiently large finite sets and form Markov chains as:

$$U - S - (V,T,X_1,X_2,X_3),$$
$$V - T - (U,S,X_1,X_2,X_3),$$
$$S - X_1 - (X_2,X_3),$$
$$T - X_2 - (X_1,X_3)$$

In addition, the following bound is an explicit upper bound which can be easily deduced from Theorem 1 of [1]:

$$R_1 \leq I(X_1; X_3 | X_2)$$
$$R_2 \leq I(X_2; X_3 | X_1)$$

The proof is given in Section V. B.

*Corollary 1*: If user 2 is only interested in wiretapping and not sharing a secret key with user 3, random variables $T$ and $V$ can be assumed to be constant. In this case, the lower bound of Theorem 1 coincides with the upper bound of Theorem 2 and the forward secret key capacity between the users 1 and 3 would be equal to:

$$R_1 = \max\{I(S; X_3 | U) - I(S; X_2 | U)\}$$

for random variables $U, S$ which form a Markov chain as $U - S - X_1 - X_2, X_3$. This result is in agreement with the result of Theorem 1 of [1].

*Theorem 3 (inner bound of the backward key capacity region)*: In the backward secret key strategy of the described source model, the rate pair $(R_1, R_2)$ is an achievable key rate pair if:

$$R_1 \geq 0, R_2 \geq 0$$
$$R_1 \leq \{I(S; X_1 | U) - I(S; X_2, T | U)\}$$
$$R_2 \leq \{I(T; X_2 | U) - I(T; X_1, S | U)\}$$

where $U$, $S$ and $T$ are random variables taking values in sufficiently large finite sets and according to the distribution:

$$p(u, s, t, x_1, x_2, x_3) = p(u | s, t) p(s, t | x_3) p(x_1, x_2, x_3).$$

The proof is given in Section V. C. Intuitive interpretation of Theorem 3 is as follows. In the case of backward key capacity region, only user 3 is permitted to send information to users 1 and 2. In this case, user 3 considers two random variables $S, T$ with distribution $p(s, t | x_3)$ and intends to send required information so that users 1 and 2 can reconstruct random variables $S$ and $T$, respectively, and then user 3 exploits these random variables for sharing secret keys with these users. First, it transmits realizations of random variable $U$ which has distribution $p(u | s, t)$ and then sends information with rate $H(S | X_1, U)$ so that user 1 can reconstruct $S$ and information with rate $H(T | X_2, U)$ so that user 2 can reconstruct $T$. Consequently, user 2 has access to random variables $X_2, U, T$ and also information with rate $H(S | X_1, U)$ for obtaining information about user 1's key. So:

$$R_1 = H(S) - I(S; X_2, U, T) - H(S | U, X_1) = I(S; U, X_1) - I(S; X_2, U, T) = I(S; X_1 | U) - I(S; X_2, T | U)$$

With the same approach the rate $R_2$ can be deduced.

***Theorem 4 (outer bound of the backward key capacity region)***: In the backward secret key strategy of the described source model, if the rate pair $(R_1, R_2)$ is an achievable key rate pair, then it satisfies:

$$R_1 \geq 0, R_2 \geq 0$$
$$R_1 \leq \min\{I(S;X_1|U) - I(S;X_2|U), I(S;X_1|T,U) - I(S;X_2|T,U)\}$$
$$R_2 \leq \min\{I(T;X_2|U) - I(T;X_1|U), I(T;X_2|S,U) - I(T;X_1|S,U)\}$$

where $U$, $S$ and $T$ are random variables taking values in sufficiently large finite sets and according to the distribution $p(u,s,t,x_1,x_2,x_3) = p(u|s,t)p(s,t|x_3)p(x_1,x_2,x_3)$ which form Markov chains as $U - S - X_3$ and $U - T - X_3$.

In addition, the following bound is an explicit upper bound which can be easily deduced from Theorem 1 of [1]:

$$R_1 \leq I(X_1; X_3|X_2)$$
$$R_2 \leq I(X_2; X_3|X_1)$$

The proof is given in Section V. D.

***Corollary 2***: If user 2 is only interested in wiretapping and not sharing a secret key, the random variable $T$ can be assumed to be constant. In this case, the lower bound of Theorem 3 coincides with the upper bound of Theorem 4 and the backward secret key capacity between user 1 and 3 would be equal to:

$$R_1 = \max\{I(S;X_1|U) - I(S;X_2|U)\}$$

for the random variables which form Markov chain as $U - S - X_3 - X_1, X_2$. This result is in agreement with the result of Theorem 1 of [1].

## IV. SPECIAL CASES

In his section, we discuss some special cases in which the secret key capacity region can be found.

***Case 1***: When sources $X_1, X_2$ and $X_3$ form a Markov chain as $X_1 - X_2 - X_3$, then the forward and backward key capacity regions reduce to:

$$R_1 = 0$$
$$0 \leq R_2 \leq I(X_2; X_3|X_1)$$

The achievability is obtained by replacing $S = X_1, T = X_2, U = V = \phi$ in Theorem 1 and $T = X_3, S = U = \phi$ in Theorem 3. It should be noted that because of the above Markov chain, the equality $I(X_2; X_3) - I(X_1; X_3) = I(X_2; X_3|X_1)$ holds. For the converse part of the forward and backward key capacity regions, we directly exploit Theorems 2 and 4, respectively.

When sources $X_1, X_2$ and $X_3$ form a Markov chain as $X_2 - X_1 - X_3$, the secret key capacity region can be derived by symmetry from case 1.

***Case 2***: When sources $X_1, X_2$ and $X_3$ form a Markov chain as $X_1 - X_3 - X_2$, then the forward key capacity region reduces to:

$$0 \leq R_1 \leq I(X_1; X_3 | X_2)$$
$$0 \leq R_2 \leq I(X_2; X_3 | X_1)$$

The achievability is obtained by replacing $S = X_1, T = X_2, U = V = \phi$ in Theorem 1. It should be noted that because of the above Markov chain, the equalities $I(X_1; X_3) - I(X_2; X_1) = I(X_1; X_3 | X_2)$ and $I(X_2; X_3) - I(X_2; X_1) = I(X_2; X_3 | X_1)$ hold. The converse part can be directly followed from Theorem 2.

***Case 3***: When sources $X_1, X_2$ and $X_3$ form a Markov chain as $X_1 - X_3 - X_2$, then the backward key capacity region reduces to:

$$R_1 \geq 0, R_2 \geq 0$$
$$R_1 \leq I(S; X_1 | U) - I(S; X_2 | U)$$
$$R_2 \leq I(T; X_2 | U) - I(T; X_1 | U)$$

where $U$, $S$ and $T$ are random variables taking values in sufficiently large finite sets and according to the distribution $p(u, s, t, x_1, x_2, x_3) = p(u|s,t) p(s,t|x_3) p(x_1, x_2, x_3)$ which form Markov chains as:

$$U - S - X_3$$
$$U - T - X_3$$
$$S - X_1 - X_2 - T$$

The existence of such random variables $S$ and $T$ can be deduced from the Markov chain $X_1 - X_3 - X_2$. This situation is shown in Fig.3. For these random variables, we have $I(S;T|X_1,U) = I(S;T|X_2,U) = 0$ and so, achievability can be deduced from Theorem 3. The converse part can be directly deduced from Theorem 2.

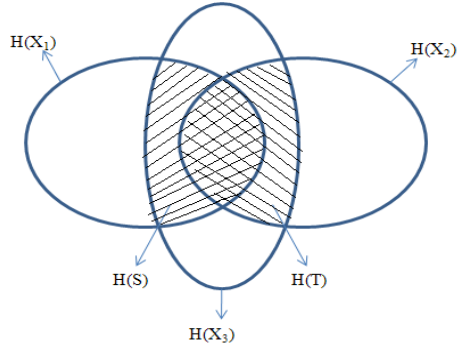

Fig.3. An example for the case $X_1 - X_3 - X_2$

## V. PROOFS

In this section, proofs of the theorems in Section III are given.

### A. PROOF OF THEOREM 1

**Construction of the Codebooks**

First, we describe random codebook generation at users 1 and 2. For a distribution $p(s)$, collection of codewords $s^N$, each uniformly drawn from the set $T_{\varepsilon_1}^N(P_S)$, is generated by user 1. $T_{\varepsilon_1}^N(P_S)$ denotes the set of jointly typical sequences $s^N$. Similarly, for a distribution $p(t)$, collection of codewords $t^N$, each uniformly drawn from the set $T_{\varepsilon_1}^N(P_T)$, is generated by user 2. Now, for a fixed distribution $p(u|s)$, user 1 generates $2^{N(I(S;U)+\varepsilon_2)}$ i.i.d. codewords of length $N$, $U^N(a)$ for $a \in \{1,...,2^{N(I(S;U)+\varepsilon_2)}\}$ with distribution $p(u)$. Similarly, for a fixed distribution $p(v|t)$, user 2 generates $2^{N(I(T;V)+\varepsilon_2)}$ i.i.d. codewords of length $N$, $V^N(b)$ for $b \in \{1,...,2^{N(I(T;V)+\varepsilon_2)}\}$ with distribution $p(v)$.

User 1 divides the typical sequences $s^N$ into $2^{NR'_1}$ bins with the same size in a uniformly random manner where $R'_1 = (H(S|X_2,U) - R_1)$. The index of each bin is denoted as $k'$ and the corresponding random variable is denoted as $K'$. Also the codewords of each bin are randomly divided into $2^{NR_1}$ bins with the same size and the bin index of the latter bins is denoted as $k$ with the corresponding random variable $K$. It is obvious that in each internal bin with bin index $k$, there are $2^{NR''_1}$ typical sequences $s^N$ where $R''_1 = (I(S;X_2,U) + \varepsilon_1)$ which we use index $k''$ for them. Hence each typical codeword $s^N$ can be uniquely determined with three indices as $s^N_{k,k',k''}$ and vice versa. Similarly, user 2 divides the typical sequences of $t^N$ into $2^{NR'_2}$ bins with the same size in a uniformly random manner where $R'_2 = (H(T|X_1,V) - R_2)$. The bin index of each bin is denoted as $l'$ and the corresponding random variable is denoted as $L'$. Also the codewords of each bin are randomly divided into $2^{NR_2}$ bins with the same size and the bin index of the latter bins is denoted as $l$ with the corresponding random variable $L$. It is obvious that in each internal bin with bin index $l$, there are $2^{NR_2''}$ typical sequences $t^N$ where $R_2'' = I(T;X_1,V) + \varepsilon_1$ which we use index $l''$ for them. Hence each typical codeword $t^N$ can be uniquely determined with three indices as $t^N_{l,l',l''}$ and vice versa.

Now, for every typical $X_1^N = x_1^N$, all codewords $s^N$ which are jointly typical with $x_1^N$, based on distribution $p(s|x_1)$, are collected in a set which is denoted as $S^N_{x_1^N}$. In the same manner, for every typical $X_2^N = x_2^N$, all codewords $t^N$ which are jointly typical whith $x_2^N$, based on distribution $p(t|x_2)$, are collected in a set which is denoted as $T^N_{x_2^N}$. The codebooks of users 1 and 2 for $X_1^N = x_1^N$ and $X_2^N = x_2^N$ are shown in Fig.4. It is assumed that all the users are informed of the binning schemes and distributions used.

| k \ k' | 1 | 2 | • | • | • | $2^{NR_1'}$ |
|---|---|---|---|---|---|---|
| 1 | •••• | •••• | •••• | •••• | •••• | •••• |
| 2 | •••• | •••• | •••• | •••• | •••• | •••• |
| • | •••• | •••• | •••• | •••• | •••• | •••• |
| • | •••• | •••• | •••• | •••• | •••• | •••• |
| • | •••• | •••• | •••• | •••• | •••• | •••• |
| $2^{NR_1}$ | •••• | •••• | •••• | •••• | •••• | •••• |

$S^N_{x_1^N}$ : Set of user 1's codewords for $X_1^N = x_1^N$

| l \ l' | 1 | 2 | • | • | • | $2^{NR_2'}$ |
|---|---|---|---|---|---|---|
| 1 | •••• | •••• | •••• | •••• | •••• | •••• |
| 2 | •••• | •••• | •••• | •••• | •••• | •••• |
| • | •••• | •••• | •••• | •••• | •••• | •••• |
| • | •••• | •••• | •••• | •••• | •••• | •••• |
| • | •••• | •••• | •••• | •••• | •••• | •••• |
| $2^{NR_2}$ | •••• | •••• | •••• | •••• | •••• | •••• |

$T^N_{x_2^N}$ : Set of user 2's codewords for $X_2^N = x_2^N$

Fig.4. Codebooks of users 1 and 2 for $X_1^N = x_1^N$ and $X_2^N = x_2^N$

**Encoding**

For encoding, users 1 and 2 observe the i.i.d. sequences $X_1^N$ and $X_2^N$, e.g., $x_1^N$ and $x_2^N$, respectively, and select the corresponding sets $S_{x_1^N}^N$ and $T_{x_2^N}^N$. User 1 randomly selects a sequence $s^N$ from the set $S_{x_1^N}^N$ and chooses the respective row index ($k$) of the codeword (as shown in Fig.4) as secret key with user 3 and sends the respective column index ($k'$) of the codeword over the public channel. He also sends index $a$ of its jointly typical sequence $U^N(a)$ over the public channel. Similarly, user 2 randomly selects a sequence $t^N$ from the set $T_{x_2^N}^N$ and chooses the respective row index ($l$) of the codeword (as shown in Fig.4) as secret key with user 3 and sends the respective column index ($l'$) of the codeword over the public channel. He also sends the index $b$ of its jointly typical sequence $V^N(b)$ over the public channel.

**Decoding and Error Probability Analysis**

For decoding, user 3 receives the indices $k', a, l', b$ from the public channel and also observes the i.i.d. sequences $X_3^N$ e.g., $x_3^N$. User 3 decodes the pair $(s_{k,k',k''}^N, t_{l,l',l''}^N)$ if:

$$(s_{k,k',k''}^N, t_{l,l',l''}^N, x_3^N) \in T_{\varepsilon_0}^{(N)}(P_{S,T,X_3|U,V})$$

when such pair $(s_{k,k',k''}^N, t_{l,l',l''}^N)$ exists and is unique and otherwise, he declares error. After decoding such $(s_{k,k',k''}^N, t_{l,l',l''}^N)$, user 3 chooses the indices $k$ and $l$ as secret keys with users 1 and 2, respectively.

Now, we analyze the decoding probability of error. Without loss of generality, we assume that the codewords $s_{1,1,1}^N$ and $t_{1,1,1}^N$ are, respectively, chosen by users 1 and 2 and so the key pair $(s_{1,1,1}^N, t_{1,1,1}^N)$ should be decoded by user 3. The event $E$ is defined as:

$$E(k,k',k'',l,l',l'') \triangleq \{(s_{k,k',k''}^N, t_{l,l',l''}^N, x_3^N) \in T_{\varepsilon_0}^{(N)}(P_{S,T,X_3|U,V})\}$$

The decoding error probability is bounded as:

$$P_e^{(N)} \leq P\{E^c(1,1,1,1,1,1) \mid (s_{1,1,1}^N, t_{1,1,1}^N) \text{chosen}\} + \sum_{(k,k'') \neq (1,1), (l,l'') \neq (1,1)} P\{E(k,1,k'',l,1,l'') \mid (s_{1,1,1}^N, t_{1,1,1}^N) \text{chosen}\}$$
$$+ \sum_{(k,k'') \neq (1,1)} P\{E(k,1,k'',1,1,1) \mid (s_{1,1,1}^N, t_{1,1,1}^N) \text{chosen}\} + \sum_{(l,l'') \neq (1,1)} P\{E(1,1,1,l,1,l'') \mid (s_{1,1,1}^N, t_{1,1,1}^N) \text{chosen}\}$$

The first term vanishes due to the joint asymptotic equipartition property (AEP):

$$P\{E^c(1,1,1,1,1,1) \mid (k,l) = (1,1) \text{ sent}\} \leq \varepsilon_0$$

In the second term for $(k,k'') \neq (1,1)$ and $(l,l'') \neq (1,1)$ we have (according to the Slepian-Wolf Theorem [12])

$$P\{E(k,1,k'',l,1,l'')|(k,l) = (1,1)\text{ sent}\} \leq 2^{-R_1'-R_2'+N(H(S,T|X_3,U,V)+\varepsilon_0)}$$

In the third term for $(k,k'') \neq (1,1)$ we have:

$$P\{E(k,1,k'',1,1,1)|(k,l) = (1,1)\text{ sent}\} \leq 2^{-R_1'+N(H(S|X_3,T,U,V)+\varepsilon_0)} = 2^{-R_1'+N(H(S|X_3,T,U)+\varepsilon_0)}$$

Finally, in the forth term for $(l,l'') \neq (1,1)$ we have:

$$P\{E(1,1,1,l,1,l'')|(k,l) = (1,1)\text{ sent}\} \leq 2^{-R_2'+N(H(T|X_3,S,U,V)+\varepsilon_0)} = 2^{-R_2'+N(H(T|X_3,S,V)+\varepsilon_0)}$$

and hence, the decoding error probability can be bounded as:

$$P_e^{(N)} \leq \varepsilon_0 + 2^{-N(R_1'+R_2'-H(S,T|X_3,U,V)-\varepsilon_0)} + 2^{-N(R_1'-H(S|X_3,T,U)-\varepsilon_0)} + 2^{-N(R_2'-H(T|X_3,S,V)-\varepsilon_0)}$$

If we set:

$$R_1' \geq H(S|X_3,T,U)$$
$$R_2' \geq H(T|X_3,S,V)$$
$$R_1' + R_2' \geq H(S,T|X_3,U,V)$$

or in other words:

$$R_1 \leq H(S|X_2,U) - H(S|X_3,T,U) \overset{(a)}{=} H(S|X_2,T,U) - H(S|X_3,T,U) = I(S;X_3|T,U) - I(S;X_2|T,U)$$
$$R_2 \leq H(T|X_1,V) - H(T|X_3,S,V) \overset{(b)}{=} H(T|X_1,S,V) - H(T|X_3,S,V) = I(T;X_3|S,V) - I(T;X_1|S,V)$$
$$R_1 + R_2 \leq H(S|X_2,U) + H(T|X_1,V) - H(S,T|X_3,U,V) = I(S,T;X_3|U,V) - I(S;X_2|T,U) - I(T;X_1|S,V) - I(S;T|U,V)$$

then for any $\varepsilon_0 > 0$, $P_e^{(N)} \leq 4\varepsilon_0$ and if we set $4\varepsilon_0 = \varepsilon$, then the reliability condition 5 in Definition 1 will be satisfied. It should be noted that in the above equations, equalities (a) and (b) follow from the distributions of random variable $S$ and $T$. It is obvious that the encoding scheme can satisfy the uniformity conditions (9) and (10) in Definition 1.

**Analysis of Security Conditions**

Now, we should analyze the security conditions (6) and (7) in Definition 1. User 2 attempts to obtain information about user 1's key and to this end, he exploits $M_2, X_2^N$ and the information which is sent by user 1 on the public channel:

$$I(K;M_2,X_2^N,K',U^N) \overset{(a)}{=} I(K;X_2^N,K',U^N) = H(K) - H(K|X_2^N,K',U^N)$$

$$= H(K) - H(K,S^N|X_2^N,K',U^N) + H(S^N|K,X_2^N,K',U^N)$$

$$= H(K) - H(K|X_2^N,K',U^N,S^N) - H(S^N|X_2^N,K',U^N) + H(S^N|K,X_2^N,K',U^N)$$

$$\overset{(b)}{=} H(K) - H(S^N|X_2^N,K',U^N) + H(S^N|K,X_2^N,K',U^N)$$

$$= NH(S|X_2,U) - NR_1' - H(S^N|X_2^N,K',U^N) + H(S^N|K,X_2^N,K',U^N)$$

$$\overset{(c)}{\leq} H(S^N|X_2^N,U^N) + N\varepsilon_1' - NR_1' - H(S^N|X_2^N,K',U^N) + H(S^N|K,X_2^N,K',U^N)$$

$$= I(S^N;K'|X_2^N,U^N) - NR_1' + H(S^N|K,X_2^N,K',U^N) + N\varepsilon_1'$$

$$= H(K'|X_2^N,U^N) - H(K'|S^N,X_2^N,U^N) - NR_1' + H(S^N|K,X_2^N,K',U^N) + N\varepsilon_1'$$

$$\leq H(K') - H(K'|S^N,X_2^N,U^N) - NR_1' + H(S^N|K,X_2^N,K',U^N) + N\varepsilon_1'$$

$$\overset{(d)}{=} H(K') - NR_1' + H(S^N|K,X_2^N,K',U^N) + N\varepsilon_1'$$

$$= H(S^N|K,X_2^N,K',U^N) + N\varepsilon_1'$$

$$\overset{(e)}{\leq} N(\varepsilon_1' + \varepsilon_2')$$

In the above equations, (a) follows from the independence of $M_2$ from other random variables, (b) from the fact that the index $k$ is one of the indices of $s^N$ and the equality $H(K|X_2^N,K',U^N,S^N) = 0$ holds. For proving (c), we use Lemma 1 which is given in part A of the Appendix. Equality (d) is true because the index $k'$ is also one of the indices of $s^N$. Finally for (e), we use Lemma 2 (which is given in part B of the Appendix) to show that:

$$H(S^N|K,X_2^N,K',U^N) \leq N\varepsilon_2'.$$

Similarly, the security condition for user 2's key is satisfied as:

$$I(L;M_1,X_1^N,L',V^N) \leq N(\varepsilon_3' + \varepsilon_4')$$

and so, the security conditions (6) and (7) of Definition 1 are satisfied when $\varepsilon_i' = \frac{1}{2}\varepsilon, i = 1,2,3,4$.

## B. PROOF OF THEOREM 2

For deriving upper bound of the forward key capacity region, we use the reliable and secure transmission conditions. In the forward key strategy, users 1 and 2, respectively, generate the keys $K$ and $L$ for sharing with user 3:

$$K = K(M_1,X_1^N), L = L(M_2,X_2^N)$$

Then, users 1 and 2, respectively, generate $F_1$ and $F_2$ where $F_1 = f_1(M_1,X_1^N), F_2 = f_2(M_2,X_2^N)$ and transmit them over the public channel so that user 3 can reconstruct $K$ and $L$ with an arbitrary probability of error $\varepsilon > 0$. According to Fano's inequality:

$$\frac{1}{N}H(K,L|M_3,X_3^N,F_1,F_2) \leq H(\varepsilon) + \varepsilon(\log|\mathcal{K}||\mathcal{L}|-1) \triangleq \varepsilon_1$$

After reconstructing these keys, user 3 uses $K$ and $L$ as secret keys with users 1 and 2, respectively, and for arbitrarily small $\varepsilon > 0$, the following security conditions should be satisfied:

$$I(K;M_2,X_2^N,F_1) \leq N.\varepsilon,$$
$$I(L;M_1,X_1^N,F_2) \leq N.\varepsilon$$

Now, we show that for keys that satisfy the reliability and security conditions described above, there exist random variables $U, V, S, T$ that form Markov chains as mentioned in Theorem 2 and satisfy the following relations:

$$H(K) \leq I(S;T,X_3|U) - I(S;X_2|U) + \varepsilon'$$
$$H(L) \leq I(T;S,X_3|V) - I(T;X_1|V) + \varepsilon'$$

We prove upper bound for $R_1$. The proof for $R_2$ can be deduced by symmetry.

$$\frac{1}{N}H(K) \stackrel{(a)}{\leq} \frac{1}{N}H(K|M_2,X_2^N,F_1) + \varepsilon$$
$$\stackrel{(b)}{\leq} \frac{1}{N}H(K|M_2,X_2^N,F_1) - \frac{1}{N}H(K|M_3,X_3^N,F_1,F_2,L) + \varepsilon_1 + \varepsilon$$
$$\stackrel{(c)}{=} \frac{1}{N}H(K|X_2^N,F_1) - \frac{1}{N}H(K|X_3^N,F_1,F_2,L) + \varepsilon_1 + \varepsilon$$
$$= \frac{1}{N}[I(K;X_3^N,F_2,L|F_1) - I(K;X_2^N|F_1)] + \varepsilon_1 + \varepsilon$$
$$\stackrel{(d)}{\leq} \frac{1}{N}\sum_{i=1}^{N}[I(K;X_{3,i},F_2,L|X_{3,1}^{i-1},X_{2,i+1}^N,F_1) - I(K;X_{2,i}|X_{3,1}^{i-1},X_{2,i+1}^N,F_1)] + \varepsilon_1 + \varepsilon$$
$$\stackrel{(e)}{\leq} \frac{1}{N}\sum_{i=1}^{N}[I(S_i;T_i,X_{3,i}|U_i) - I(S_i;X_{2,i}|U_i)] + \varepsilon_1 + \varepsilon$$
$$\stackrel{(f)}{\leq} [I(S_Q;T_Q,X_{3,Q}|U_Q) - I(S_Q;X_{2,Q}|U_Q)] + \varepsilon'$$

where (a) results from the security condition, (b) from Fano's inequality, (c) from independence of $(M_2, M_3)$ from other random variables, (d) from Lemma 3 (which is given in part C of the Appendix), (e) from definition of the random variables $U, V, S, T$ as:

$$U_i = (X_{3,1}^{i-1}, X_{2,i+1}^N, F_1), V_i = (X_{3,1}^{i-1}, X_{1,i+1}^N, F_2), S_i = (K,U_i), T_i = (L,V_i)$$

and (f) from definition of the random variable $Q$ which is uniformly distributed on $\{1,2,...,N\}$ and setting $\varepsilon' = \varepsilon_1 + \varepsilon$.

Similarly, by using the above mentioned variables we have:

$$R_2 = \frac{1}{N}H(L) \leq [I(T_Q;S_Q,X_{3,Q}|V_Q) - I(T_Q;X_{1,Q}|V_Q)] + \varepsilon'$$

It can be seen that the desired equations are satisfied with random variables which form Markov chains as in Theorem 2.

### C. PROOF OF THEOREM 3

**Construction of the Codebooks**

First, we describe random codebook generation at user 3. For a distribution $p(s,t)$, collection of codewords, $(s^N,t^N)$ each uniformly drawn from the set $T_{\varepsilon_1}^N(P_{S,T})$, is generated by user 3. Now, for a fixed distribution $p(u|s,t)$, user 3 generates $2^{N(I(S,T;U)+\varepsilon_2)}$ i.i.d. codewords of length $N$, $U^N(a)$ for $a \in \{1,...,2^{N(I(S,T;U)+\varepsilon_2)}\}$ with distribution $p(u)$.

User 3 divides the typical sequences of $s^N$ into $2^{NR_1'}$ bins with the same size in a uniformly random manner where $R_1' = (H(S|X_2,T,U) - R_1)$. The bin index of each bin is denoted as $k'$ and the corresponding random variable is denoted as $K'$. Also the codewords of each bin are randomly divided into $2^{NR_1}$ bins with the same size and the bin index of the latter bins is denoted as $k$ with the corresponding random variable $K$. It is obvious that in each internal bin with bin index $k$, there are $2^{NR_1''}$ typical sequences $s^N$ where $R_1'' = (I(S;X_2,T,U) + \varepsilon_1)$ which we use index $k''$ for them. Hence each typical codeword $s^N$ can be uniquely determined with three indices as $s^N_{k,k',k''}$ and vice versa. Also, user 3 divides typical sequences of $t^N$ into $2^{NR_2'}$ bins with the same size in a uniformly random manner where $R_2' = (H(T|X_1,S,U) - R_2)$. The bin index of each bin is denoted as $l'$ and the corresponding random variable is denoted as $L'$. Also the codewords of each bin are randomly divided into $2^{NR_2}$ bins with the same size and the bin index of the latter bins is denoted as $l$ with the corresponding random variable $L$. It is obvious that in each internal bin with bin index $l$, there are $2^{NR_2''}$ typical sequences $t^N$ where $R_2'' = (I(T;X_1,S,U) + \varepsilon_1)$ which we use index $l''$ for them. Hence each typical codeword $t^N$ can be uniquely determined with three indices as $t^N_{l,l',l''}$ and vice versa.

Now, for every typical $X_3^N = x_3^N$, all codewords $(s^N,t^N)$ which are jointly typical with $x_3^N$, based on distribution $p(s,t|x_3)$, are collected in a set which is denoted as $(S^N,T^N)_{x_3^N}$. It is assumed that all the users are informed of the binning schemes and distributions used.

**Encoding**

For encoding, user 3 observes the i.i.d. sequence of $X_3^N$ e.g., $x_3^N$ and after selecting the corresponding set $(S^N,T^N)_{x_3^N}$,

he randomly selects a sequence $(s^N, t^N)$ from this set. Then, he chooses the respective row index ($k$) of the codeword $s^N$ (as shown in Fig.4) as secret key with user 1 and sends the respective column index ($k'$) of the codeword over the public channel. Also, he chooses the respective row index ($l$) of the codeword $t^N$ (as shown in Fig.4) as secret key with user 2 and sends the respective column index ($l'$) of the codeword over the public channel. In addition, user 3 sends index $a$ of $U^N(a)$ which is jointly typical with the sequence $(s^N, t^N)$ over the public channel.

**Decoding and Error Probability Analysis**

For decoding, users 1 and 2 receive the indices $k', l', a$ from the public channel and also observe the i.i.d. sequences $X_1^N$ and $X_2^N$ e.g., $x_1^N$ and $x_2^N$, respectively. User 1 decodes $s_{k,k',k''}^N$ if:

$$(s_{k,k',k''}^N, x_1^N) \in T_{\varepsilon_0}^{(N)}(P_{S,X_1|U})$$

when such $s_{k,k',k''}^N$ exists and is unique and otherwise he declares error. User 2 decodes $t_{l,l',l''}^N$ if:

$$(t_{l,l',l''}^N, x_2^N) \in T_{\varepsilon_0}^{(N)}(P_{T,X_2|U})$$

when such $t_{l,l',l''}^N$ exists and is unique and otherwise he declares error.

Now we analyze decoding error probability. We define:

$$P_e^{(N)} = \max\{P_{e1}^{(N)}, P_{e2}^{(N)}\}$$

where $P_{e1}^{(N)}$ and $P_{e2}^{(N)}$ are, respectively, decoding error probabilities at users 1 and 2. Without loss of generality, we assume that the codewords $s_{1,1,1}^N$ and $t_{1,1,1}^N$ are chosen by user 3 and so, $s_{1,1,1}^N$ and $t_{1,1,1}^N$ should be decoded by users 1 and 2, respectively. Events $E_1$ and $E_2$ are defined as:

$$E_1(k, k', k'') \triangleq \{(s_{k,k',k''}^N, x_1^N) \in T_{\varepsilon_0}^{(N)}(P_{S,X_1|U})\}$$

$$E_2(l, l', l'') \triangleq \{(t_{l,l',l''}^N, x_2^N) \in T_{\varepsilon_0}^{(N)}(P_{T,X_2|U})\}$$

The decoding error probabilities are bounded as:

$$P_{e1}^{(N)} \leq P\{E_1^c(1,1,1) | (s_{1,1,1}^N, t_{1,1,1}^N) \text{chosen}\} + \sum_{(k,k'') \neq (1,1)} P\{E(k,1,k'') | (s_{1,1,1}^N, t_{1,1,1}^N) \text{chosen}\}$$

$$P_{e2}^{(N)} \leq P\{E_2^c(1,1,1) | (s_{1,1,1}^N, t_{1,1,1}^N) \text{chosen}\} + \sum_{(l,l'') \neq (1,1)} P\{E(l,1,l'') | (s_{1,1,1}^N, t_{1,1,1}^N) \text{chosen}\}$$

According to the joint asymptotic equipartition property (AEP), decoding error probabilities can be bounded as:

$$P_{e1}^{(N)} \leq \varepsilon_0 + 2^{-R_1' + N(H(S|X_1,U) + \varepsilon_0)}$$
$$P_{e2}^{(N)} \leq \varepsilon_0 + 2^{-R_2' + N(H(T|X_2,U) + \varepsilon_0)}$$

and if we set:

$$R_1' \geq H(S|X_1,U)$$
$$R_2' \geq H(T|X_2,U)$$

or in other words:

$$R_1 \leq H(S|X_2,T,U) - H(S|X_1,U) = I(S;X_1|U) - I(S;X_2,T|U)$$
$$R_2 \leq H(T|X_1,S,U) - H(T|X_2,V) = I(T;X_2|U) - I(T;X_1,S|U)$$

then for any $\varepsilon_0 > 0$, $P_{ei}^{(N)} \leq 2\varepsilon_0$ for $i = 1,2$ and so $P_e^{(N)} \leq 2\varepsilon_0$ and if we set $2\varepsilon_0 = \varepsilon$, then the reliability condition 5 in Definition 1 will be satisfied. It is obvious that the encoding scheme can satisfy the uniformity conditions (9) and (10) in Definition 1.

**Analysis of Security Conditions**

Now, we should analyze the security conditions (6) and (7) in Definition 1. User 2 attempts to obtain information about user 1's key and to this end, he exploits $M_2, X_2^N$ and the information which is sent by user 3 over the public channel, i.e., the indices $k'$, $l'$ and $a$:

$$I(K; M_2, X_2^N, K', L', U^N) \stackrel{(a)}{=} I(K; X_2^N, K', L', U^N) \leq I(K; X_2^N, K', L', T^N, U^N) = I(K; X_2^N, K', T^N, U^N) + I(K; L'|X_2^N, K', T^N, U^N)$$

$$\stackrel{(b)}{=} I(K; X_2^N, K', T^N, U^N) = H(K) - H(K|X_2^N, K', T^N, U^N)$$

$$= H(K) - H(K, S^N | X_2^N, K', T^N, U^N) + H(S^N | K, X_2^N, K', T^N, U^N)$$

$$= H(K) - H(K | X_2^N, K', T^N, U^N, S^N) - H(S^N | X_2^N, K', T^N, U^N) + H(S^N | K, X_2^N, K', T^N, U^N)$$

$$\stackrel{(c)}{=} H(K) - H(S^N | X_2^N, K', T^N, U^N) + H(S^N | K, X_2^N, K', T^N, U^N)$$

$$= NH(S|X_2,T,U) - NR_1' - H(S^N | X_2^N, K', T^N, U^N) + H(S^N | K, X_2^N, K', T^N, U^N)$$

$$\stackrel{(d)}{\leq} H(S^N | X_2^N, T^N, U^N) + N\varepsilon_1' - NR_1' - H(S^N | X_2^N, K', T^N, U^N) + H(S^N | K, X_2^N, K', T^N, U^N)$$

$$= I(S^N; K' | X_2^N, T^N, U^N) - NR_1' + H(S^N | K, X_2^N, K', T^N, U^N) + N\varepsilon_1'$$

$$= H(K' | X_2^N, T^N, U^N) - H(K' | S^N, X_2^N, T^N, U^N) - NR_1' + H(S^N | K, X_2^N, K', T^N, U^N) + N\varepsilon_1'$$

$$\leq H(K') - H(K' | S^N, X_2^N, T^N, U^N) - NR_1' + H(S^N | K, X_2^N, K', T^N, U^N) + N\varepsilon_1'$$

$$\stackrel{(e)}{=} H(K') - NR_1' + H(S^N | K, X_2^N, K', T^N, U^N) + N\varepsilon_1'$$

$$= H(S^N | K, X_2^N, K', T^N, U^N) + N\varepsilon_1'$$

$$\stackrel{(f)}{\leq} N(\varepsilon_1' + \varepsilon_2')$$

In above equations, (a) follows from the independence of $M_2$ from other random variables, (b) from the fact that given $T^N, L'$ is impendent of other random variables, (c) from the fact that the index $k$ is one of the indices of $s^N$ and the equality $H(K|X_2^N, K', T^N, U^N, S^N) = 0$ holds. For proving (d), we use the same approach as in Lemma 1 which is given in part A of the Appendix. Equality (e) is true because the index $k'$ is also one of the indices of $s^N$. Finally for (f), we use the same approach as in Lemma 2 (which is given in part B of the Appendix) to show that:

$$H(S^N | K, X_2^N, K', T^N, U^N) \leq N\varepsilon_2'.$$

Similarly, the security condition for user 2's key is satisfied as:

$$I(L; M_1, X_1^N, K', L', U^N) \leq N(\varepsilon_3' + \varepsilon_4')$$

and so, the security conditions (6) and (7) of Definition 1 are satisfied when $\varepsilon_i' = \frac{1}{2}\varepsilon, i = 1, 2, 3, 4$.

### D. PROOF OF THEOREM 4

For deriving upper bounds of the backward key capacity region, we use the reliable and secure transmission conditions. In the backward key strategy, user 3 generates the keys $K$ and $L$ for sharing with users 1 and 2, respectively:

$$K = K(M_3, X_3^N), L = L(M_3, X_3^N)$$

Also, it sends $F_3$ over the public channel where $F_3 = f_3(M_3, X_3^N)$ to enable users 1 and 2 to compute $K$ and $L$, respectively, with an arbitrary probability of error $\varepsilon > 0$. According to Fano's inequality:

$$\frac{1}{N} H(K | M_1, X_1^N, F_3) \leq H(\varepsilon) + \varepsilon(\log|\mathcal{K}| - 1) \triangleq \varepsilon_1,$$
$$\frac{1}{N} H(L | M_2, X_2^N, F_3) \leq H(\varepsilon) + \varepsilon(\log|\mathcal{L}| - 1) \triangleq \varepsilon_2$$

Also the security conditions require that:

$$I(K; M_2, X_2^N, F_3) \leq N\varepsilon,$$
$$I(L; M_1, X_1^N, F_3) \leq N\varepsilon$$

Now, we derive upper bounds for $R_1$. The proofs for $R_2$ can be deduced by symmetry. For the first upper bound of $R_1$:

$$\frac{1}{N}H(K) \overset{(a)}{\leq} \frac{1}{N}H(K|M_2, X_2^N, F_3) + \varepsilon$$

$$\overset{(b)}{\leq} \frac{1}{N}[H(K|M_2, X_2^N, F_3) - H(K|M_1, X_1^N, F_3)] + \varepsilon_1 + \varepsilon$$

$$\overset{(c)}{=} \frac{1}{N}[H(K|X_2^N, F_3) - H(K|X_1^N, F_3)] + \varepsilon_1 + \varepsilon$$

$$= \frac{1}{N}[I(K; X_1^N|F_3) - I(K; X_2^N|F_3)] + \varepsilon_1 + \varepsilon$$

$$\overset{(d)}{=} \frac{1}{N}\sum_{i=1}^{N}[I(K; X_{1,i}|X_1^{i-1}, X_{2,i+1}^N, F_3) - I(K; X_{2,i}|X_1^{i-1}, X_{2,i+1}^N, F_3)] + \varepsilon_1 + \varepsilon$$

$$\overset{(e)}{=} \frac{1}{N}\sum_{i=1}^{N}[I(S_i; X_{1,i}|U_i)] - I(S_i; X_{2,i}|U_i)] + \varepsilon_1 + \varepsilon$$

$$\overset{(f)}{=} [I(S_Q; X_{1,Q}|U_Q) - I(S_Q; X_{2,Q}|U_Q)] + \varepsilon'$$

where (a) results from the security condition, (b) from Fano's inequality at user 1, (c) from independence of $(M_1, M_2)$ from other random variables, (d) from Lemma 3 (in which the random variable $F_2$ is set to be constant), (e) from definition of the random variables $U, S, T$ as:

$U_i = (X_1^{i-1}, X_{2,i+1}^N, F_3)$, $S_i = (K, U_i)$, $T_i = (L, U_i)$ and (f) from definition of the random variable $Q$ which is uniformly distributed on $\{1, 2, ..., N\}$ and setting $\varepsilon' = \varepsilon_1 + \varepsilon$.

For the second upper bound of $R_1$, we have:

$$\frac{1}{N}H(K) \overset{(a)}{\leq} \frac{1}{N}H(K|M_2, X_2^N, F_3) + \varepsilon \leq \frac{1}{N}H(K, L|M_2, X_2^N, F_3) + \varepsilon$$

$$= \frac{1}{N}H(K|M_2, X_2^N, F_3, L) + \frac{1}{N}H(L|M_2, X_2^N, F_3) + \varepsilon$$

$$\overset{(b)}{\leq} \frac{1}{N}H(K|M_2, X_2^N, F_3, L) + \varepsilon_2 + \varepsilon$$

$$\overset{(c)}{\leq} \frac{1}{N}H(K|M_2, X_2^N, F_3, L) - H(K|M_1, X_1^N, F_3)] + \varepsilon_1 + \varepsilon_2 + \varepsilon$$

$$\overset{(d)}{=} \frac{1}{N}H(K|X_2^N, F_3, L) - H(K|X_1^N, F_3)] + \varepsilon_1 + \varepsilon_2 + \varepsilon$$

$$\leq \frac{1}{N}H(K|X_2^N, F_3, L) - H(K|X_1^N, F_3, L)] + \varepsilon_1 + \varepsilon_2 + \varepsilon$$

$$= \frac{1}{N}[I(K; X_1^N|F_3, L) - I(K; X_2^N|F_3, L)] + \varepsilon_1 + \varepsilon_2 + \varepsilon$$

$$\overset{(e)}{=} \frac{1}{N}\sum_{i=1}^{N}[I(K; X_{1,i}|X_1^{i-1}, X_{2,i+1}^N, F_3, L) - I(K; X_{2,i}|X_1^{i-1}, X_{2,i+1}^N, F_3, L)] + \varepsilon_1 + \varepsilon_2 + \varepsilon$$

$$\overset{(f)}{=} \frac{1}{N}\sum_{i=1}^{N}[I(S_i; X_{1,i}|U_i, T_i)] - I(S_i; X_{2,i}|U_i, T_i)] + \varepsilon_1 + \varepsilon_2 + \varepsilon$$

$$\overset{(g)}{=} [I(S_Q; X_{1,Q}|U_Q, T_Q) - I(S_Q; X_{2,Q}|U_Q, T_Q)] + \varepsilon''$$

where (a) results from the security condition, (b) from Fano's inequality at user 2, (c) from Fano's inequality at user 1, (d) from independence of $(M_1, M_2)$ from other random variables, (e) from Lemma 3 (in which the random variable $F_2$ is set to be constant), (f) from definition of the random variables $U, S, T$ as above and (g) from definition of the random variable $Q$ as above and setting $\varepsilon'' = \varepsilon_1 + \varepsilon_2 + \varepsilon$.

Following the same approach, upper bounds for $R_2$ can be deduced and so Theorem 4 is proved for some random variables with distribution $p(u,s,t,x_1,x_2,x_3) = p(u|s,t)p(s,t|x_3)p(x_1,x_2,x_3)$ which form Markov chains as $U - S - X_3$ and $U - T - X_3$.

## VI. CONCLUSIONS

In this paper, a source model for secret key generation were studied in which each of users 1 and 2 intends to share a secret key with user 3 where user 1 acts as a wiretapper for user 2 and vice versa. Three users could observe i.i.d outputs of correlated sources and there is a public channel between users. In the described model, the forward and backward key strategies were considered based on the direction of the public channel, i.e., from users 1 and 2 to user 3 or in the reverse direction. For both the forward and backward key strategies, inner and outer bounds of secret key capacity regions were derived. Our results also include the results of previous works such as [1]. Our upper and lower bounds did not coincide generally but some special cases were considered where these bounds were tight.

As the continuation of this work, we are now exploring a model similar to the described model but instead of the public channel, there is a generalized multiple access channel (MAC) between the terminals, where users 1 and 2 govern the inputs of the MAC and outputs are received by users 1, 2 and 3. Also as the future works, we can suggest the same problem of this paper for the situation where there is a two-way public channel i.e., from users 1 and 2 to user 3 and vice versa. Also unlimited usage of the public channel can be viewed as a generalization of the problem.

## APPENDIX

### A. LEMMA 1

For sufficiently large $N$ and sufficiently small $\varepsilon_1'$, we have:

$$NH(S|X_2,U) \leq H(S^N|X_2^N,U^N) + N\varepsilon_1'$$

Proof: We use the indicator function:

$$\mu(s,x_2,u) = \begin{cases} 1, & (s^N, x_2^N, u^N) \notin A_{\varepsilon_0'}^{(N)}(P_{S,X_2,U}) \\ 0, & \text{otherwise} \end{cases}$$

We have:

$$I(S^N; X_2^N, U^N) \leq I(S^N, \mu; X_2^N, U^N)$$

and hence:

$$H(S^N | X_2^N, U^N) \geq H(S^N) - I(S^N, \mu; X_2^N, U^N) = NH(S) - I(S^N, \mu; X_2^N, U^N)$$
$$= NH(S) - I(S^N; X_2^N, U^N | \mu) - I(\mu; X_2^N, U^N)$$
$$= NH(S) - P(\mu=1)I(S^N; X_2^N, U^N | \mu=1) - P(\mu=0)I(S^N; X_2^N, U^N | \mu=0) - I(\mu; X_2^N, U^N)$$

We analyze the above terms one by one.

For the second term:

$$P(\mu=1)I(S^N; X_2^N, U^N | \mu=1) \leq NP[(s^N, x_2^N, u^N) \notin A_{\varepsilon_0'}^{(N)}(P_{S,X_2,U})] \log|\mathcal{S}| \leq N\varepsilon_0' \log|\mathcal{S}|$$

For the third term:

$$P(\mu=0)I(S^N; X_2^N, U^N | \mu=0) \leq I(S^N; X_2^N, U^N | \mu=0)$$
$$= \sum_{(s^N, x_2^N, u^N) \in A_{\varepsilon_0'}^{(N)}(P_{S,X_2,U})} P(s^N, x_2^N, u^N)[\log P(s^N, x_2^N, u^N) - \log P(s^N) - \log P(x_2^N, u^N)]$$
$$\leq N(H(S) + H(X_2, U) - H(S, X_2, U) + 3\varepsilon_0') = N(I(S; X_2, U) + 3\varepsilon_0')$$

For the forth term:

$$I(\mu; X_2^N, U^N) \leq H(\mu) \leq 1$$

Finally, we can deduce:

$$H(S^N | X_2^N, U^N) \geq NH(S) - N\varepsilon_0' \log|\mathcal{S}| - N(I(S; X_2, U) + 3\varepsilon_0') - 1 = N(H(S | X_2, U) - \underbrace{(\varepsilon_0' \log|\mathcal{S}| + 3\varepsilon_0' + \frac{1}{N})}_{\varepsilon_1'})$$

## B. Lemma 2

For sufficiently large $N$ and sufficiently small $\varepsilon_2'$, in the forward key strategy, we have:

$$H(S^N | K, X_2^N, K', U^N) \leq N\varepsilon_2'$$

Proof: For fixed $k$ and $k'$, we assume that user 1 transmits a codeword $s_{k,k',k''}^N$ where $1 \leq k \leq 2^{NR_1}$, $1 \leq k' \leq 2^{NR_1'}$ and $1 \leq k'' \leq 2^{NR_1''}$. First, we show that user 2 can decode $s_{k,k',k''}^N$ with sufficiently small probability of error if it has access to sequences $k, k', x_2^N, u^N$. User 2 selects $k''$ so that:

$$(s_{k,k',k''}^N, x_2^N, u) \in A_{\varepsilon_3'}^{(N)}(P_{S,X_2,U})$$

if such $k''$ exists and it is unique, otherwise we declare error. With the assumption that $s_{k,k',1}^N$ is sent by user 1, error occurred when $(s_{k,k',1}^N, x_2^N, u) \notin A_{\varepsilon_3'}^{(N)}(P_{S,X_2,U})$ or when $(s_{k,k',k''}^N, x_2^N, u) \in A_{\varepsilon_3'}^{(N)}(P_{S,X_2,U})$ for $k'' \neq 1$. Due to joint AEP:

$$P((s_{k,k',1}^N, x_2^N, u) \notin A_{\varepsilon_3'}^{(N)}(P_{S,X_2,U})) \leq \varepsilon_3'$$

and also:

$$P_{k'' \neq 1}\{(s_{k,k',k''}^N, x_2^N, u) \in A_{\varepsilon_3'}^{(N)}(P_{S,X_2,U})\} \leq 2^{NR_1'' - N(I(S;X_2,U - \varepsilon_3'))} = 2^{N(\varepsilon_1 + \varepsilon_3')}$$

So, we can bound decoding error of user 2 as:

$$P_e \leq \varepsilon_3' + 2^{N(\varepsilon_1 + \varepsilon_3')}$$

and by choosing $\max\{\varepsilon_1, \varepsilon_3'\} \to 0$, we can make $P_e$ sufficiently small.

Now, we exploit Fano's inequality to obtain:

$$\frac{1}{N} H(S^N | K, X_2^N, K', U^N) \leq \frac{1}{N}[1 + P_e R_1''] \leq \frac{1}{N} + \frac{1}{N}(\varepsilon_3' + 2^{N[\varepsilon_1 + \varepsilon_3']})[I(S;X_2,U) + \varepsilon_1] \triangleq \varepsilon_2'$$

### C. LEMMA 3

This lemma is a modified version of Lemma 1 in [1].

For arbitrary random variables $K, F_1, F_2$ and sequences of random variables $X_2^N, X_3^N$ we have [1]:

$$[I(K; X_3^N, F_2 | F_1) - I(K; X_2^N | F_1)] \leq \sum_{i=1}^{N} [I(K; F_2, X_{3,i} | X_{3,1}^{i-1}, X_{2,i+1}^N, F_1) - I(K; X_{2,i} | X_{3,1}^{i-1}, X_{2,i+1}^N, F_1)]$$

Proof: First, we consider the right hand side of the above inequality:

$$\sum_{i=1}^{N} [I(K; F_2, X_{3,i} | X_{3,1}^{i-1}, X_{2,i+1}^N, F_1) - I(K; X_{2,i} | X_{3,1}^{i-1}, X_{2,i+1}^N, F_1)] = \sum_{i=1}^{N} [H(K | X_{3,1}^{i-1}, X_{2,i}^N, F_1) - H(K | X_{3,1}^i, X_{2,i+1}^N, F_1, F_2)]$$

$$= H(K | X_{2,1}^N, F_1) - \sum_{i=1}^{N-1} [H(K | X_{3,1}^i, X_{2,i+1}^N, F_1) - H(K | X_{3,1}^i, X_{2,i+1}^N, F_1, F_2)] - H(K | X_{3,1}^N, F_1, F_2)$$

$$= I(K; F_2, X_{3,1}^N | F_1) - I(K; X_{2,1}^N | F_1) + \underbrace{\sum_{i=1}^{N-1} [I(K; F_2 | X_{3,1}^i, X_{2,i+1}^N, F_1)]}_{\geq 0} \geq I(K; F_2, X_{3,1}^N | F_1) - I(K; X_{2,1}^N | F_1)$$

# REFERENCES

[1] R. Ahlswede and I. Csiszár, "Common randomness in information theory and cryptography, part I: Secret sharing," *IEEE Trans. Inf. Theory*, vol. 39, no. 4, pp. 1121–1132, Jul. 1993.

[2] I. Csiszár and J. Körner, "Broadcast channels with confidential messages," *IEEE Trans. Inf. Theory*, vol. 24, no. 3, pp. 339–348, May 1978.

[3] I. Csiszár and P. Narayan, "Secrecy capacities for multiple terminals," *IEEE Trans. Inf. Theory*, vol. 50, no. 12, pp. 3047–3061, Dec. 2004.

[4] I. Csiszár and P. Narayan, "Secrecy capacities for multiterminal channel model," *IEEE Trans. Inf. Theory*, vol. 54, no. 6, pp. 2437–2452, Jun. 2008.

[5] A. A. Gohari and V. Anantharam, "New bounds on the information-theoretic key agreement of multiple terminals", in *Proc. IEEE Int. Symp. Inf. Theory (ISIT)*, Toronto, Canada, pp. 742-746, Jul. 2008.

[6] A. A. Gohari and V. Anantharam, "Information-theoretic key agreement of multiple terminals - Part I: Source model", *IEEE Trans. Inf. Theory, submitted*, Jun. 2008.

[7] A. A. Gohari and V. Anantharam, "Information-theoretic key agreement of multiple terminals - Part II: Channel model", *IEEE Trans. Inf. Theory, submitted*, Jun. 2008.

[8] U. M. Maurer, "Secret key agreement by public discussion from common information," *IEEE Trans. Inf. Theory*, vol. 39, no. 3, pp. 733–742, May 1993.

[9] U. Maurer and S. Wolf, "Information-theoretic key agreement: From weak to strong secrecy for free," in *Proc. EUROCRYPT'2000, LNCS*, vol. 1807, Bruges, Belgium: Springer-Verlag, pp. 351–368, May 2000.

[10] R. Renner and S. Wolf, "New bounds in secret-key agreement: the gap between formation and secrecy extraction," in *Proc. EUROCRYPT'03, LNCS*, Warsaw, Poland: Springer-Verlag, pp. 562-577, May 2003.

[11] C. E. Shannon, "Communication theory of secrecy systems," *AT&T Bell Labs. Tech. J.*, vol. 28, pp. 656–715, 1949.

[12] D. Slepian and J. K. Wolf, "Noiseless coding of correlated information sources," *IEEE Trans. Inf. Theory*, vol. 19, no. 4, pp. 471–480, Jul. 1973.

[13] A. Wyner, "The wire-tap channel," *AT&T Bell Labs. Tech. J.*, vol. 54, pp. 1355–1387, 1975.

[14] C. Ye and P. Narayan, "The private key capacity region for three terminals," in *Proc. IEEE Int. Symp. Inf. Theory (ISIT)*, Chicago, USA, p. 44, Jun. 2004.

[15] C. Ye and A. Rezenik, "Group secret key generation algorithms," in *Proc. IEEE Int. Symp. Inf. Theory (ISIT)*, Nice, France, pp. 2596-2600, Jun. 2007.